\documentclass[twocolumn,nofootinbib,showpacs,prd,superscriptaddress]{revtex4}

\usepackage{amsmath}
\usepackage{amsfonts}
\usepackage{amssymb}
\usepackage{graphicx}

\usepackage{color}

\begin{document}

\title{Cosmological dark energy effects from entanglement }

\author{Salvatore Capozziello}
\affiliation{Dipartimento di Fisica, Universit\`a di Napoli ''Federico II'', Via Cinthia, 80126 Napoli, Italy.}
\affiliation{Istituto Nazionale di Fisica Nucleare (INFN), Sez. di Napoli, Via Cinthia, Napoli,  80126  Italy.}

\author{Orlando Luongo}
\affiliation{Dipartimento di Fisica, Universit\`a di Napoli ''Federico II'', Via Cinthia, 80126 Napoli, Italy.}
\affiliation{Istituto Nazionale di Fisica Nucleare (INFN), Sez. di Napoli, Via Cinthia, Napoli,  80126 Italy.}
\affiliation{Instituto de Ciencias Nucleares, Universidad Nacional Autonoma de M\'exico (UNAM), Mexico.}

\author{Stefano Mancini}
\affiliation{Scuola di Scienze \& Tecnologie, Universit\`a di Camerino, 62032 Camerino, Italy}
\affiliation{ Istituto Nazionale di Fisica Nucleare (INFN), Sez. di Perugia, Via Pascoli, 06123 Perugia, Italy.}

\begin{abstract}
The thorny issue of relating  information theory  to cosmology is here  addressed by  assuming  a possible connection between quantum entanglement measures and observable universe. In particular, we
propose a cosmological toy model, where the equation of state of the cosmological fluid, which drives the today observed cosmic acceleration, can be inferred from quantum  entanglement between different cosmological epochs. In such a way  the dynamical dark energy results as byproduct of quantum entanglement.
\end{abstract}

\pacs{98.80.-k, 03.67.-a}

\date{\today}

\maketitle

\section{introduction}

The problem of quantizing gravity is still an open question of modern physics. In particular, despite  experimental success in the  weak field limit, the Einstein theory fails to be predictive at high energy regimes, i.e. at UV scales \cite{bingo,117www,Hartle:1986gn}. Unfortunately, no observational evidences are able to probe  quantum  gravity regime. Then, various phenomenological approaches are trying to describe quantum effects of general relativity. Examples of recent developments have been carried out by postulating Lorentz invariance violation \cite{Horawa,Orla,edue,BlackHoles} or by considering possible signatures in   the  Wheeler DeWitt equation   \cite{GeneralizzazioneHorawa,adddreee, Muinpark,noirev}. On the other hand, the basic idea of \emph{quantum cosmology}  is to find out quantum predictions at cosmological scales. To this regard, in the framework of a homogeneous and isotropic universe, such a problem can be simplified assuming the so-called {\it minisuperspace approach} (see \cite{odirev} for a recent review on the argument). However, it is likely that the existence of such quantum cosmological effects could be even  unrelated to quantum gravity. In other words, many authors put forward strong arguments that quantum cosmology could represent an independent branch for studying quantum effects in general relativity \cite{indep}.
At the same time, it has been argued that quantum cosmology might be related to quantum information theory
\cite{Terno04} and the notion of \emph{quantum entanglement} can play a role in learning about curved space times \cite{ivette} or cosmological features \cite{tremendo}.

Following this line, we present here an approach to quantum cosmology that relies on \emph{entangled cosmological states}. The underlying idea consists in choosing a multipartite quantum system where each party corresponds to an epoch of the universe.
Then, we investigate whether the equation of state (EoS) of the cosmological fluid \cite{rev}, which drives the today observed  cosmic acceleration, could be inferred from the above quantum picture.
 This allows us to express the present dark energy density by
the amount of entanglement existing
between cosmological epochs. In particular, we show that a toy model, based on entanglement of two cosmological epochs, is able to describe the dynamics related to cosmic acceleration \cite{v1,v2,v3,lambda1,lambda2,lambda3,lambda4}.
To this end, we consider the so called {\it negativity} as measure of entanglement \cite{negativity}.
This opens up the way to express quantum parameters, derived from negativity, in terms of  observable quantities, such as mass density, curvature and so forth. Moreover, the dynamical effects of dark energy could be inferred by studying the entanglement measure of such entangled cosmological states. In particular, we find from considerations on negativity, an evolving EoS. In addition, we show that the physical properties of such an EoS are able to reproduce the universe dynamics at late times \cite{coppa,bousso}.

The paper is organized as follows. In sec. II, we discuss the role of entanglement negativity in cosmology. In Sec. III, we describe in detail the corresponding cosmological model, derived from negativity and we give particular attention to the consequences in the observable universe. Finally, in Sec. IV, we deal with conclusions and future perspective of our work.

%%%%%%%%%%%%%%%%%%%%%%%%%%%%%%%%%%%%%%%%%%%%

\section{ Entangled cosmological states}

 Let us start taking into account  the minimal number of cosmological observables assuming the hypothesis of homogeneity and isotropy (that is the cosmological principle) and we use the Friedmann-Robertson-Walker (FRW) metric, supposing the existence of further terms expressed as barotropic fluid within the Friedmann equations \cite{lindo}. To clarify this statement, let us denote by $P_{QE}$ and $\rho_{QE}$ the  pressure and the density associated to the quantum effects respectively; then the Friedmann equations for the Hubble parameter
 $H$, function of (cosmological) time read \cite{coppa}
\begin{eqnarray}
  H^2 &=& \frac{8\pi G}{3} \rho -k(1+z)^2\,, \label{F1}\\
  \dot{H} +H^2 &=& -\frac{4\pi G}{3}\left(3 P+{ \rho}\right)\,,\label{F2}
\end{eqnarray}
 where the dot is the derivative with respect to the cosmic time and  we defined $\rho=\rho_{m}+\rho_{QE}$ and $P=P_{m}+P_{QE}$, with $\rho_m$ and $P_m$ respectively the density and pressure of matter.
Furthermore, $k$ denotes the scalar curvature of the universe and $z$ the redshift. Notice that the following relation, between the total density $\rho=P/w$ and the Hubble parameter $H$, holds true
\begin{equation}
\rho=\frac{3}{8\pi G}H^2.
\label{rhoH}
\end{equation}

Each quantum state of the universe leads to the existence of a quantum fluid, described by an EoS of the form
\begin{equation}
w_{QE}\equiv\frac{P_{QE}}{\rho_{QE}}\,.
\end{equation}
The presence of such a fluid may correspond to the existence of dark energy.

Given that, we consider two cosmological observable quantities,
 the densities of the cosmological fluid $\Omega_{m}$ and the scalar curvature $\Omega_{k}$,
it is natural to specify a state of the universe in each epoch by a two dimensional vector.
To lie in a quantum framework we consider such a vector over the complex field $\mathbb{C}$, i.e.
of the form
\begin{equation}
 |\phi\rangle\equiv \left(
\Omega_{m}+i\Omega_{k},\\
\Omega_{k}+i\Omega_{m}\right)^T,
\label{phi}
\end{equation}
 thus belonging to the Hilbert space $\mathbb{C}^2$.

Looking at the form of $|\phi\rangle$ in \eqref{phi} we can take as linear independent
(non-normalized) vectors in $\mathbb{C}^2$ the following ones
\begin{eqnarray}
|\tilde e_A\rangle&=&\left(
\Omega_{m}+i\Omega_{k},
i\Omega_{m}+\Omega_{k}\right)^T, \\
|\tilde e_B\rangle&=&\left(
\Omega_{m}-i\Omega_{k},
-i\Omega_{m}+\Omega_{k}\right)^T.
\end{eqnarray}
Then, the Gram-Schmidt procedure leads to the following orthonormal basis
\begin{eqnarray}
|e_A\rangle&=&
N_A |\tilde e_A\rangle,\label{ortho1}\\
|e_B\rangle&=&
N_B (\Omega_m^2+\Omega_k^2) |\tilde e_B\rangle
+N_B 2i\Omega_m\Omega_k |\tilde e_A\rangle,
\label{ortho2}
\end{eqnarray}
where
\begin{eqnarray}
N_A&=&\frac{1}{\sqrt{2(\Omega_m^2+\Omega_k^2)}},\\
N_B&=&\frac{1}{\sqrt{2(\Omega_m^2+\Omega_k^2)(\Omega_k^2-\Omega_m^2)^2}}.
\end{eqnarray}

%%%%%%%%%%%%%%%%%%%%%%%%%%%%%%%%%%%%%%%%%%%%%%%%%%%

The above described quantities and dynamics \eqref{F1}, \eqref{F2} refer to a given epoch. Now, in order to account for different epochs we are going to consider different Hilbert spaces.
The idea of associating to each epoch of the universe a different Hilbert space arises in order to satisfy the basic demands of the standard cosmological model \cite{berny1,berny2,berny3}. In particular, it is clear, from the Friedmann equations, that different epochs are characterized by different dynamical properties. Examples of epochs are in fact inflation, reheating, recombination, and so forth. Unfortunately, the only redshift dynamical evolution is not enough to guarantee the kinematical effects of separate epochs, as the universe expands \cite{berny4}. This is probably due to the different interactions between cosmological species such as radiations, baryons, cold dark matter, etc. Thus, to depict the physical changes among various epochs, the standard cosmological model assumes the existence of phase transitions \cite{berny5}. No evidences for such phase transitions are however measured, so, for our purposes, all the phase transitions can be replaced by entanglement processes among different epochs. In doing so, each epoch is mostly characterized by a certain Hilbert space.

%%%%%%%%%%%%%%%%%%%%%%%%%%%%%%%%%%%%%%%%%%%%%%%%%%%

 Then, the simplest multipartite system involves two epochs and the associated Hilbert space is
$\mathbb{C}^2\otimes \mathbb{C}^2$ with an orthonormal basis obtainable
by tensoring vectors of the kind of \eqref{ortho1}, \eqref{ortho2}
\begin{eqnarray}
 |e_A\rangle_{1} |e_A\rangle_{2},\;
  |e_A\rangle_{1} |e_B\rangle_{2},\;
   |e_B\rangle_{1} |e_A\rangle_{2},\;
    |e_B\rangle_{1} |e_B\rangle_{2},
 \label{basis}
\end{eqnarray}
where the subscripts $1,2$ define the epochs of interest.

Our \emph{entangled states ansatz} (ESA) is to consider the universe in an entangled state between the two epochs, e.g.  a state of the form
\begin{equation}
|\Psi\rangle = \alpha|e_A\rangle_1|e_B\rangle_2 + \beta|e_B\rangle_1|e_A\rangle_2\,,
\label{stato}
\end{equation}
where, $\alpha,\beta\in\mathbb{C}$ such that
\begin{equation}
|\alpha|^2+|\beta|^2=1.
\label{norm}
\end{equation}

Here, we are  interested in investigating the entanglement properties  in order to measure their effects  on cosmological observables. In particular, our aim is to find out a relation between a quantum entanglement measure  with the evolution of cosmological densities $\Omega_{m}$, $\Omega_k$  in one epoch.

We consider the  {\it negativity} as a measure of entanglement \cite{negativity}, that is
\begin{equation}\label{ns}
\mathcal{N}=2\sum_k \max(0,-\lambda_k)\,,
\end{equation}
where the sum is over the eigenvalues of the partially transposed density matrix.

Looking at \eqref{stato} the density matrix $|\Psi\rangle\langle\Psi|$ in the basis \eqref{basis} reads
\begin{eqnarray}
\left(
\begin{array}{cccc}
0 & 0 & 0 & 0\\
0 & |\alpha|^2 & \alpha\beta^* & 0\\
0 & \alpha^*\beta & |\beta|^2 & 0\\
0 & 0 & 0 & 0
\end{array}
\right).
\end{eqnarray}
Then, its partial transpose with respect to the first system becomes
\begin{eqnarray}
\left(
\begin{array}{cccc}
0 & 0 & 0 & \alpha\beta^*\\
0 & |\alpha|^2 & 0 & 0\\
0 & 0 & |\beta|^2 & 0\\
\alpha^*\beta & 0 & 0 & 0
\end{array}
\right),
\end{eqnarray}
whose eigenvalues are
\begin{equation}
  \lambda_1 = |\alpha|^2,\;
  \lambda_2 = |\beta|^2,\;
  \lambda_3 = |\alpha\beta|,\;
  \lambda_4 = -|\alpha\beta|.
  \label{evalT1}
\end{equation}
Thus, using \eqref{evalT1} in \eqref{ns}, it results
\begin{equation}
\mathcal{N}=2|\alpha\beta|.
\label{negpsi}
\end{equation}

%%%%%%%%%%%%%%%%%%%%%%%%%%%%%%%%%%%%%%%%%%%%%

\section{Negativity in Cosmology}

In order to infer the entanglement effects in the observable universe, we are now interested in relating entanglement measure like negativity to dark energy density  in one epoch.
To this end, the negativity can be reasonably assumed to be a function of $t$. Actually,
shifting hereafter from $t$ to $z$,
 we can expand it in power series  of the inverse of the cosmological scale factor $a\equiv\frac{1}{1+z}$ so that, to the lowest non-zero order, we have
\begin{equation}\label{nocomment}
\mathcal{N}(z)= N_0+N_1(1+z)^2\,.
\end{equation}
with $N_0,N_1\ge 0$.
We discard the linear term $\propto a^{-1}$, because a cosmological fluid corresponding to it does not have a clear physical meaning \cite{antro}.

 We notice that Eq. (\ref{nocomment}) represents the physical information that is held in the universe, when one considers the ansatz (\ref{phi}). The use of Eq. (\ref{nocomment}) holds inside the universe horizon, whose size scales as $R_H\propto H^{-1}$. Inside the universe horizon, it is easy to guarantee that two or more epoches, throughout the universe evolution, may be entangled to each other. In particular, our choice leads to a first approximation of a more complicated expansion of $\mathcal{N}(z)$ in powers of $R_H$.

 On the other hand, we remark the fact that, by choosing Eq. ($\ref{nocomment}$), we avoid singularities of $\mathcal{N}$, in the redshift limit $z\ll1$. However, it turns out that possible divergences could occur in the more general case $\mathcal{N}\propto R_{H}^{-n}$, where  $n$ is an integer number, depending on the "entanglement degree" between cosmic states. For our purposes, we point out that this is not the case of late time cosmology, which provides a sharp dependence on time of cosmological observables. It follows that Eq. (\ref{nocomment}) behaves smoothly in our regime, showing no phase transitions between the epoches we are dealing with.

From Eq. \eqref{negpsi} we get $|\alpha|=\frac{\mathcal{N}}{2|\beta|}$ and then using \eqref{nocomment} we obtain
\begin{equation}
|\alpha|=\frac{N_0+N_1(1+z)^2}{2|\beta|}.
\label{modal}
\end{equation}

Now let us focus on one epoch and consider quantities referred to it. Then, the involved densities
respect the triangle equality extended to
  $z\ge 0$ \cite{kumba1,kumba2}, namely
\begin{equation}
\Omega_m(z)+\Omega_k(z)+\Omega_X(z)=1,
\label{triangle}
\end{equation}
where $\Omega_X$ denotes the dark energy density.

At this point we can equate the l.h.s. of \eqref{triangle} to the l.h.s. of \eqref{norm}
\begin{equation}
\Omega_m+\Omega_k+\Omega_X=|\alpha|^2+|\beta|^2\,.
\label{1}
\end{equation}
Using \eqref{modal} into \eqref{1},
and taking into account  that
 $\Omega_k(z)\approx 0$ and $\Omega_m(z)=\omega_m (1+z)^3$
with $\omega_m$ a constant \cite{kumba3},
we can derive $\Omega_X(z)$ as
\begin{equation}\label{OmX}
\Omega_X(z)=|\beta|^2+\frac{\left(N_0+N_1(1+z)^2\right)^2}{4|\beta|^2}
- \omega_m(1+z)^3\,.
\end{equation}
The above equation relates the dark energy ($\Omega_X$) to the negativity (parameters $N_0, N_1$),
however in this relation $\beta$ is still undetermined.
In order to determine it, we can use the identity
\begin{equation}
\rho=\rho_0 \left(\omega_m (1+z)^3+\Omega_X(z)\right)
\label{rho}
\end{equation}
with $\rho_0\equiv\frac{3H_0^2}{8\pi G}$ the critical density (here $H_0=H(z=0)$).
Then, by inserting the dark energy density ($\ref{OmX}$)
into \eqref{rho} and this latter into \eqref{F1}, we arrive, using
$\Omega_k(z)\approx 0$ hence $k\approx 0$, to
\begin{equation}
H(z)=H_0\left[|\beta|^2+\frac{\left(N_0+N_1(1+z)^2\right)^2}{4|\beta|^2}\right]^{1/2}.
\label{Hubble}
\end{equation}
Finally, since $H(z=0)=H_0$ must hold true, we can use such a relation to derive $|\beta|$ as
\begin{equation}
|\beta|=\left[\frac{1+\sqrt{(1-N_0-N_1)(1+N_0+N_1)}}{2}\right]^{1/2}.
\label{beta}
\end{equation}

Now, by combining the Friedmann equations \eqref{F1}, \eqref{F2}, and using the
relation \eqref{rhoH} together with the
 fact that $\frac{dz}{dt}=-H(1+z)$, we get the continuity equation in terms of the redshift $z$
\begin{equation}
  \frac{d\rho}{dz}=3\left(\frac{1+w}{1+z}\right)\rho\,,
  \label{ce}
\end{equation}
which is equivalent to require $\nabla^\mu T_{\mu\nu}=0$, with $T_{\mu\nu}$ the energy tensor momentum of Einstein's gravity \cite{berny1}.

Dividing \eqref{ce} by $\rho_0$ we have
 \begin{equation}
  \frac{d\Omega_X}{dz}=3\left(\frac{1+w}{1+z}\right)\Omega_X\,,
\end{equation}
 and using \eqref{OmX} we get
\begin{equation}
w=\frac{-3N_0^2-2 N_0N_1(1+z)^2+N_1^2 (1+z)^4-12|\beta|^4}{3(N_0+N_1(1+z)^2)^2-12
 \omega_m(1+z)^3 |\beta|^2+12 |\beta|^4}\,.
\label{omz}
\end{equation}

Then, taking \eqref{omz} at $z=0$, and exploiting \eqref{beta}, we obtain the present time EoS
\begin{equation}
  w_0=\frac{-2N_1(N_0+N_1)+3\left(1+\sqrt{1-\left(N_0+N_1\right)^2}\right)}{3\left(
   \omega_m-1\right)\left(1+\sqrt{1-\left(N_0+N_1\right)^2}\right)}\,.
\end{equation}

Furthermore, the acceleration parameter
\begin{equation}
  q=-1+ \frac{1+z}{H}\frac{d}{dz} H\,,
\end{equation}
can be calculated by using Eqs. \eqref{Hubble} and \eqref{beta}.
At the present epoch, i.e. $z=0$, it results
\begin{equation}
  q_0=-1+\frac{N_1(N_0+N_1)}{1+\sqrt{1-\left(N_0+N_1\right)^2}}\,.
\end{equation}

Finally, the values of $w_0$ and $q_0$ known from experiments \cite{orl1,orl2,orl3,orl4}
\begin{eqnarray}
-1<w_0<0\,,
\label{valueofw0}\\
q_0>-1\,,
\label{q0ridotto}
\end{eqnarray}
can be used to bound the ranges of $N_0$ and $N_1$ as
\begin{eqnarray}
N_0\in [0,0.5],\\
 N_1\in [0,0.5]\,.
\end{eqnarray}
These bounds could give rise to two significative results. The first concerns possible measurable quantum effects today, while the second deals with the varying EoS predicted by  negativity, which seems to be able to overcome the {\it coincidence problem}, related to the standard $\Lambda$CDM picture \cite{lcdmx1}. In our case,  the cosmological constant is the zero order approximation of the dark energy term inferred from the entanglement process.

%%%%%%%%%%%%%%%%%%%%%%%%%%%%%%%%%%%%%%%%%%%%%%%%%%

\section{conclusions}

In this paper, we have investigated the possibility  that  an entanglement process between different  cosmological epochs  could give rise to dark energy effects.  In particular,  we have built up a toy model by using the hypothesis of minimal information in a FRW metric and have  taken into account different epochs of the universe evolution, identified by different space vectors (in this respect the model shares analogies with the Many World Approach to Quantum Cosmology \cite{MW}).
The corresponding cosmic dynamics reflects  the entanglement between
such spaces (epochs). Actually, the dynamical properties of entangled states are able to reproduce the connection between the dark energy density and the so-called cosmic triangle equation derived from the Friedmann equations. Thus, we inferred a relation between the properties of entanglement and the present value of the dark energy EoS, showing that it evolves in time reducing to a cosmological constant to the zeroth order (corresponding to redshift $z=0$).

Among the various  possible entangled measures \cite{Horo}, we investigated the so-called negativity,
because it leads to manageable analytical relations (this is not the case, for instance, for the von Neumann entropy of reduced density matrix). By using such a quantity, it is possible to derive a viable Hubble rate, from which naturally arise the  constraints on cosmological densities. In addition, by considering the value of negativity today, the cosmological model works fairly well with respect to the standard $\Lambda$CDM model \cite{lcdmx1}. We pointed out the possibility to compare the EoS-value and the acceleration parameter with $N_0$ and $N_1$ which give the initial conditions of negativity. Moreover, the constants $N_0$ and $N_1$ provide a way to relate the  entanglement to the observed  universe in on epoch. The result is that the considered model relates the value of the dark energy density parameter to the value of $N_1$, and this means that cosmological entanglement is, in principle, observable.

Finally, we have to  stress that what we have proposed is a simple toy model that, however, could point out a straightforward way to relate  quantum to macroscopic properties of cosmology.  More realistic models should take into account also  thermodynamical quantities and a larger set of cosmographic parameters.

\end{document}